# PAISAJE CELESTE Y ARQUEOASTRONOMÍA: LAS IGLESIAS HISTÓRICAS COMO INDICADORES DE LA PLANIMETRÍA DE UNA CIUDAD IDEAL

ALEJANDRO GANGUI

De acuerdo a la tradición popular, la ciudad canaria de San Cristóbal de La Laguna, Patrimonio de la Humanidad por la UNESCO desde 1999, surgió en los inicios del siglo XVI con una organización novedosa, acorde con un nuevo orden social pacífico inspirado por la doctrina religiosa del milenio y su expresión a través del diseño urbano. Su sistema en retícula es el primer caso conocido de ciudad colonial no fortificada con un plano regular, de todo el contexto de la expansión europea. Constituye un ejemplo prominente de lo que fue llamado una "Ciudad de Paz", el arquetipo de una ciudad-territorio que se sirvió de sus propias fronteras naturales para delimitarse. Se piensa que ese nuevo orden social surgió en el año 1500 y que habría sido alentado por los propios Reyes Católicos para la fundación de ciudades en las tierras conquistadas. La Laguna, además, sería la realización material de una ciudad ideal que fue concebida y construida según "un plan geométrico preciso". Este plan tendría un fuerte sustento en la navegación, como principal ciencia aplicada de la época, y estaría inspirado en ciertos principios filosóficos griegos, entre los cuales se hallan los lineamientos esbozados para la ciudad de Magnesia en Las Leyes de Platón.
En este trabajo discutimos brevemente algunos detalles de estas ideas, que proponen un origen geométrico planificado para la ciudad, y las fuentes originales en que se basan. Luego buscamos otros indicios que de manera más simple expliquen algunas de las características notables del ordenamiento y diseño de La Laguna, donde con el correr del tiempo se instaló un modelo de arquitectura tradicional canario que, durante el siglo XVII, se fue poblando de construcciones civiles y religiosas (iglesias, ermitas y conventos) con notables creaciones en carpintería de origen mudéjar, por ejemplo provenientes de la baja Andalucía. Para finalizar, mostramos nuestro análisis preciso de la orientación de la casi totalidad de las iglesias y ermitas históricas laguneras, tomadas en conjunto como elemento indicador de la planimetría de la ciudad. Este estudio emplea las herramientas usuales de la arqueoastronomía como complemento a la investigación histórica y cultural de las construcciones religiosas. Veremos entonces que una nueva -y más sencilla- interpretación de la distribución espacial de la ciudad es posible.

## LOS ORÍGENES DE SAN CRISTÓBAL DE LA LAGUNA

La historia colonial de la ciudad comienza en el verano boreal de 1496 cuando las tropas de la Corona de Castilla, al mando de Don Alonso Fernández de Lugo, obtienen la victoria final sobre los aborígenes de la isla de Tenerife. La zona en cuestión era estratégica: un sitio elevado y alejado del mar, con abundantes cursos de agua, rodeado por montañas y en presencia de una pequeña laguna, la que había convertido a la zona en una planicie rica y fértil.

Tras el sometimiento de los nativos, se celebra la primera fiesta del Corpus Christi en la primitiva y recién construida parroquia de Nuestra Señora de la Inmaculada Concepción, evento que corresponde a una primera fundación de la ciudad. En esa misma época se inicia el asentamiento de soldados y civiles en la antigua villa, construyéndose las primeras cabañas a los pies de la iglesia y a cierta distancia de la laguna, la que se extendía hacia el norte y que finalmente desaparecerá en 1837 cuando se le da desagüe. El núcleo inicial se conoció como la "Villa de Arriba", en donde se desarrolló un trazado irregular de calles producto de un asentamiento desordenado y sin previa parcelación.

Pero en el año 1500, Fernández de Lugo vuelve a la isla tras capitular con la Corona las condiciones oportunas en concepto de la conquista, y realiza entonces la fundación definitiva de la ciudad. Obtiene de los Reyes Católicos el título de Adelantado y la gobernación de los territorios. Como tal, ostenta el derecho pleno de administrar justicia, nombrar los diferentes cargos administrativos, judiciales y militares, adjudicar terrenos, dictar ordenanzas y ser cabeza del Cabildo, el principal y único órgano de gobierno.

La fundación definitiva de la ciudad en 1500 viene concebida a partir de un nuevo diseño en torno a la actual plaza del Adelantado, conocida entonces como "Villa de Abajo", a aproximadamente 1 km hacia el sureste de la antigua villa y alejándose de la laguna. De este nuevo núcleo surge un trazado de retícula ordenada según la distribución de calles rectas basada en el modelo clásico. Este era un patrón ya empleado en algunas ciudades de fundación anterior, por ejemplo las situadas en la baja Andalucía como Puerto Real y Santa Fe de Granada. Pero en este caso se trataba de una ciudad sin amurallar fundada bajo régimen colonial, por lo que tuvo que responder a nuevas necesidades políticas, económicas y sociales. A la larga, este nuevo diseño de ciudad serviría de modelo al proceso colonizador que posteriormente tendría lugar en los nuevos territorios americanos.

Así, la ciudad de San Cristóbal de La Laguna se fue configurando a lo largo del siglo XVI en dos núcleos de población, que surgieron, como vimos, de un modo diferenciado. El primer asentamiento provisional, elegido por el Adelantado entorno a la parroquia de la Concepción, se caracterizó por no tener un trazado urbano planificado. Tan solo unas casas de mampuesto y techumbre de paja, que dieron forma a un pequeño caserío. Unos tres años más tarde, hacia 1500, surge un segundo núcleo más racionalizado, y que promueve el asentamiento de la población al sur y al este del territorio (Serra Ràfols 1996). De hecho, el 24 de abril de ese año el Adelantado promulgó un decreto por el cual ya no se permitía la venta o la construcción de más casas en la villa antigua -incluso se prohibía la reparación de las ya existentes-, y se tomó la determinación de que las nuevas construcciones se llevaran a cabo "desde el l'espital de Santespiritus hazia el logar de Abaxo", es decir, hacia el sureste del Santo Espíritu o Convento de San Agustín (Alemán de Armas 1986, 15; Aznar Vallejo 2008, 192).

El casco histórico de la ciudad quedó prácticamente definido hacia finales del siglo XVI, como lo muestra el primer plano que se conserva (Fig. 1), dibujado por el ingeniero italiano Leonardo Torriani en 1588, quien había sido enviado por encargo del rey Felipe II para estudiar la defensa militar del archipiélago (Torriani 1978). Con los años, los dos núcleos poblacionales, las Villas de Arriba y de Abajo, terminaron uniéndose.

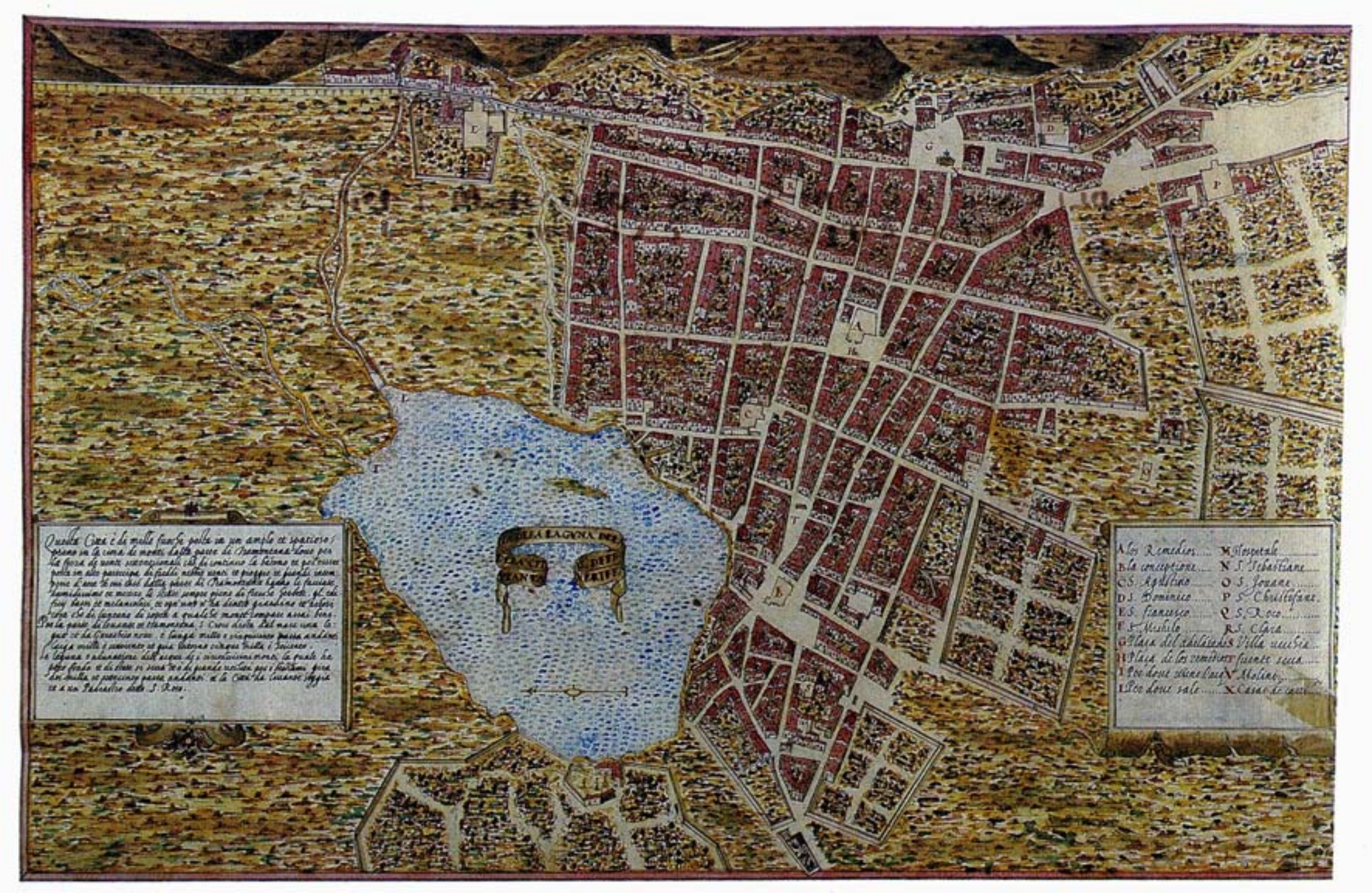

Fig. 1 Plano de la ciudad de San Cristóbal de La Laguna hacia fines del siglo XVI, dibujado por el ingeniero italiano Leonardo Torriani. Este mapa se halla "orientado", es decir que el este (o el oriente, aproximadamente) se ubica en la parte superior y, en consecuencia, el norte se halla hacia la izquierda.

Si bien en el Viejo Continente abundan las ciudades con centros históricos muy bien conservados y mucho más antiguos que el de La Laguna, lo que hace especial a esta última ciudad es que no fue el resultado de siglos de interacciones, modificaciones y reconstrucciones. San Cristóbal de La Laguna comenzó a crecer en un lugar virginal, despoblado y remoto y, desde la óptica europea, sin historia. Lo singular de la ciudad es que se cree que fue pensada como un todo desde el inicio (la ciudad como proyecto), y es ese preciso diseño original, junto a la morfología y su notable perfil urbano, el que permanece prácticamente inalterado desde los días de Torriani, luego de más de 400 años.

Sin embargo, la ciudad sería notable no sólo por lo ya descrito. Como mencionamos, La Laguna sería la realización material del primer ejemplo de ciudad no fortificada que fue concebido y construido según un plan geométrico preciso con fuerte sustento en la navegación, y cuya estructura simbólica podría interpretarse de manera similar a los mapas marinos o incluso a la distribución de las constelaciones del cielo. Su espacio habría sido organizado según un nuevo orden social pacífico surgido en el año 1500 y supuestamente impulsado por la Corona para ser implementado en la fundación de ciudades en las tierras conquistadas (Navarro Segura 1999).

En la próxima sección discutiremos brevemente algunos detalles de estas ideas que, según creemos (y hemos analizado con cierto detenimiento en Gangui y Belmonte 2017), no encuentran un adecuado sustento histórico en fuentes documentales de la época. Es por ello que uno podría genuinamente dudar de su veracidad y, en consecuencia, intentar hallar otros indicios que de manera más simple expliquen algunas de las características notables del ordenamiento y diseño de la ciudad. A continuación, y luego de hacer una reseña breve de las principales características de las construcciones religiosas de la ciudad, mostraremos nuestro análisis preciso de la orientación de la casi totalidad de las iglesias laguneras, tomadas en conjunto como elemento indicador de la planimetría de La Laguna. Veremos entonces que -por supuesto, con cierto margen de

incertidumbre- es posible hallar una nueva -y más sencilla- interpretación de la distribución espacial de la ciudad.

## ¿IDEAS GEOMÉTRICAS GRIEGAS COMO PRINCIPIO INSPIRADOR DE LA CIUDAD?

Las características de la refundación de La Laguna pueden atribuirse al propio Adelantado, resultado de su permanencia en la Corte, entre agosto y octubre de 1499, que en esos momentos residía en la ciudad de Granada. Durante esos meses, en los que Fernández de Lugo se hallaba puliendo los detalles de su futura campaña en Berbería (norte de África), se supone que la Corte dictó algunas normas sobre cómo se debería llevar adelante el proceso de la nueva fundación (Navarro Segura 1999, 204), de acuerdo con las ideas de renovación humanista y religiosa que imperaban en el momento (la doctrina del milenio ya mencionada). Del análisis comparativo entre el diseño de la ciudad y las ideas de "ciudad ideal" esbozadas en Las Leyes de Platón, se piensa que los Reyes Católicos convinieron con el Adelantado "la aplicación de los principios contenidos en la obra, como parte de un proyecto de experimentación de un nuevo modelo de ciudad a desarrollar en los nuevos territorios pacificados incorporados a la Corona de Castilla" (Navarro Segura 1999, 166).

En base a la tradición libresca de la época en la Península, hay sugerencias de que Fernández de Lugo habría tenido entre sus manos el texto de Platón, y que de su lectura minuciosa habría hecho propias las ideas del filósofo -y la posible aplicación de estas ideas en La Laguna- en lo relacionado con la ciudad de Magnesia presente en sus diálogos (Navarro Segura 2006; 1999, 186). En colaboración con Antonio de Torres, veedor de los Reyes para la campaña africana, versado en matemáticas y en navegación, y con el sevillano Pedro de Vergara, Alcalde Mayor de la ciudad lagunera durante varios años, el Adelantado habría podido llevar adelante su diseño de La Laguna en las dos primeras décadas del nuevo siglo. Sin embargo, la muerte prematura de Torres en un naufragio en 1502, y su propia desaparición en 1525, habrían relegado al olvido las ideas y el simbolismo detrás de este experimento ciudadano en los años que siguieron, durante las fundaciones de las colonias americanas (Navarro Segura 1999, 144).

Así, en el año 1500 se realiza la nueva fundación de la ciudad, que incluía la Villa de Arriba y el nuevo núcleo de población hacia el sureste impuesto por el Adelantado, y que por muchos años permanecieron físicamente separados. De acuerdo a lo ya dicho, el nuevo diseño supuestamente se apoyaba en conceptos antiguos basados en fórmulas matemáticas, y delineaba sus calles mediante el empleo de astrolabios náuticos y otros utensilios de navegación. A partir de entonces, conceptos platónicos y renacentistas comenzaron a solaparse en la configuración de un trazado urbano original. Una ciudad utópica y simbólica, proyección del cielo en la tierra, que se orientaba por la rosa de los vientos, y que tenía a la cabecera del convento de San Agustín (antiguo límite entre las dos villas) como exacto centro geométrico (Fig. 2).

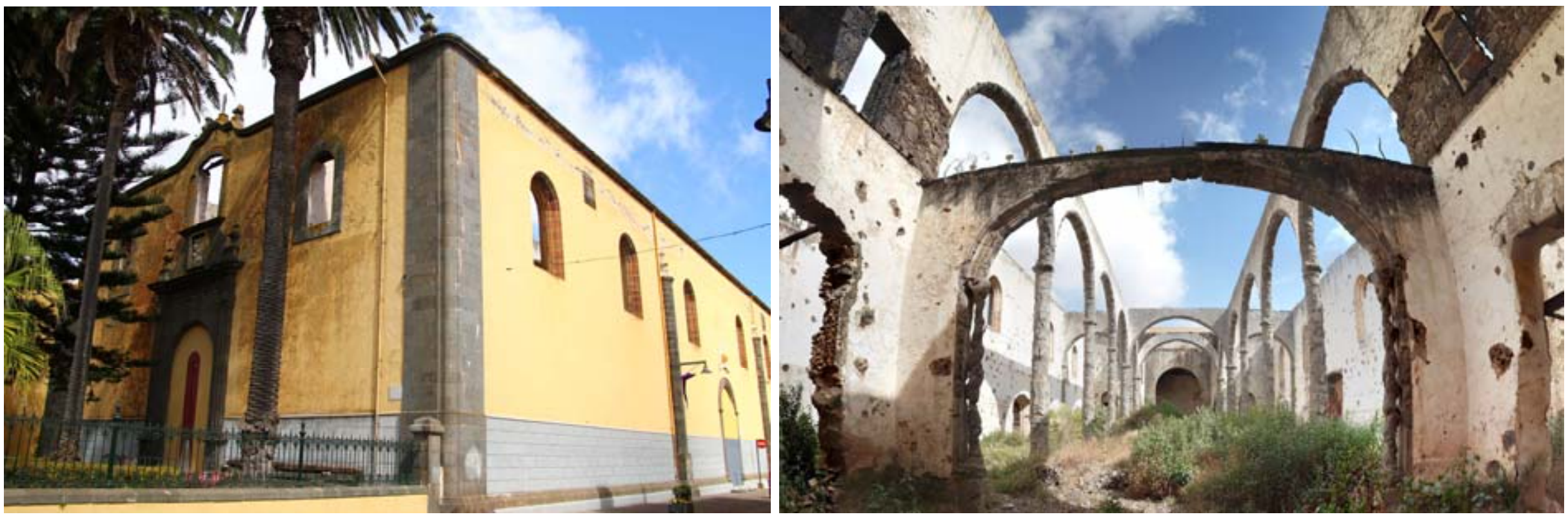

Fig. 2 Fachada principal y muros aún en pie pertenecientes a la iglesia de San Agustín, junto al antiguo convento de la misma orden, en su estado actual, luego de que el incendio de 1964 consumiera su interior e hiciera desplomarse tu techo. Su ubicación marcaba el límite suroriental de la antigua ciudad, hasta el momento en que comenzó a construirse la Villa de Abajo. Fotografías del autor.

De acuerdo al estudio de la investigadora Navarro Segura, del Departamento de Historia del Arte de la Universidad de La Laguna, publicado en su libro "La Laguna 1500", en el diseño de la ciudad se aplicaron otros conceptos geométricos adicionales, además de aquellos sugeridos en Las Leyes de Platón. En lugar prominente aparecen las instrucciones del tratado *De Architectura* de Vitruvio (c. 15 a.C.), sencillas por lo usual del empleo de instrumentos de orientación de rumbos en la navegación y el conocimiento detallado de los ocho vientos principales de la época (Navarro Segura 1999, 204). En las páginas dedicadas a la fundación de ciudades, Vitruvio relaciona los vientos como factor determinante de la salubridad de los lugares y reduce a ocho el conjunto de las regiones del horizonte. La figura que mejor se adapta a este diseño es el octógono regular inscripto en una circunferencia, y ese habría sido el empleado en La Laguna, donde el octógono vitruviano sirvió para calcular las posiciones de los ocho vientos principales y así proyectar el trazado de las calles "a medio rumbo", es decir, en direcciones intermedias a las de estos vientos, que en la zona de nuestra ciudad siempre fueron particularmente intensos.

Esta intencionalidad en el trazado geométrico de La Laguna, según la propuesta de Navarro Segura, viene reforzada por algunos ejes que atraviesan la ciudad, estableciendo equidistancias entre edificios religiosos representativos del siglo XVI. Así, el centro geométrico de San Agustín sería equidistante de cuatro templos antiguos, tomados de dos en dos: las iglesias de San Juan Bautista y del Cristo de La laguna, por un lado, y las ermitas de San Roque y de San Cristóbal, por el otro. Además, el eje de unión entre la iglesia de San Benito Abad (también del siglo XVI) y el centro del trazado en San Agustín, estaría dispuesto en forma precisa en dirección oeste-este, y dividiría el octógono ciudadano en dos partes. Por último, la distancia que separa al centro de las dos primeras iglesias sería la misma que lo separa de los dos extremos del trazado de la calle del Agua, en la actualidad calle de Nava y Grimón, calle espaldera y límite oriental de la ciudad (Fig. 3).

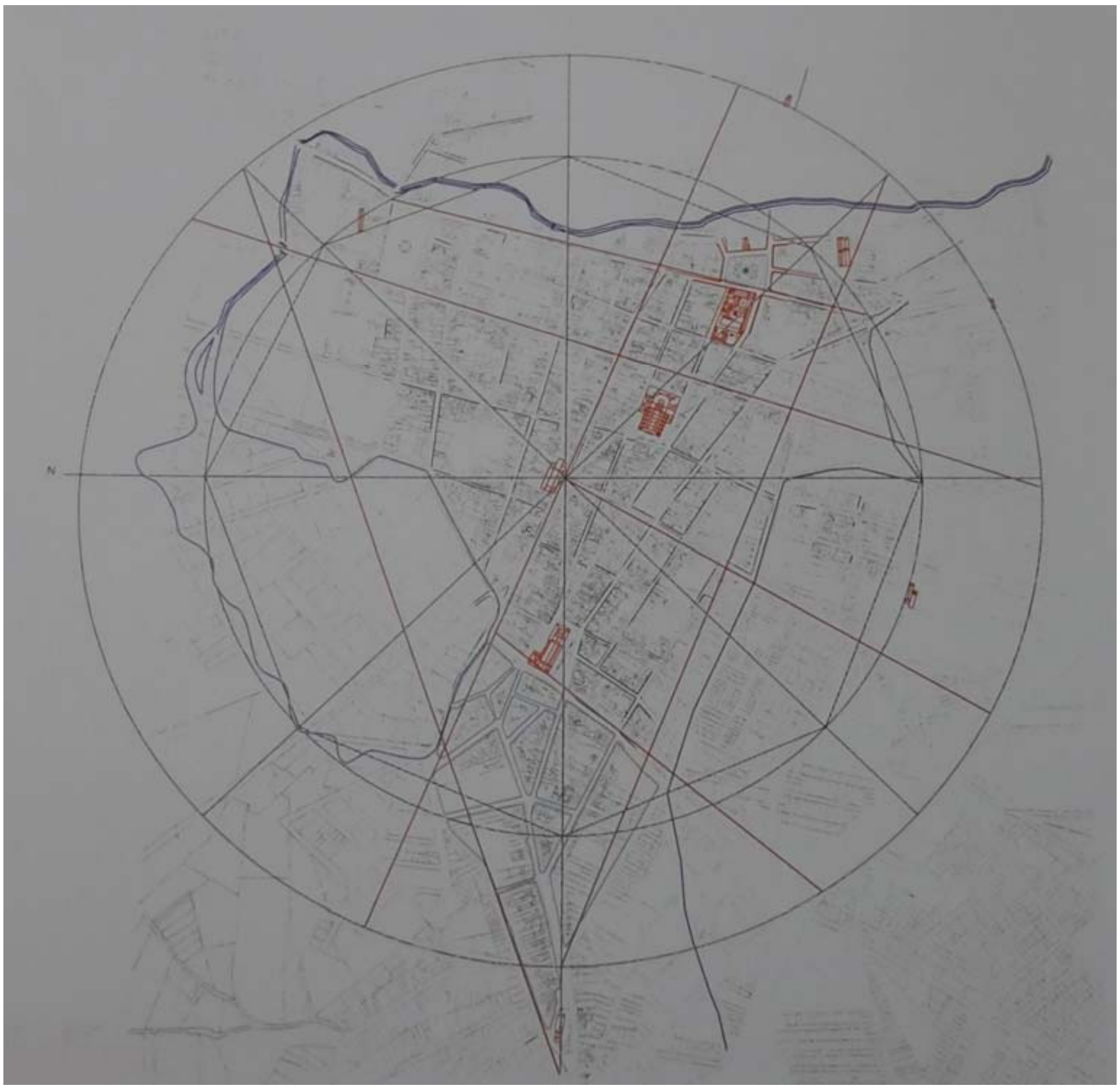

Fig. 3 Mapa dibujado por Navarro Segura con la distribución del callejero de San Cristóbal de La Laguna en un octógono vitruviano inscripto en la circunferencia y relacionado con los vientos principales. Centrado en la iglesia de San Agustín, presenta la ubicación de varios edificios religiosos como referencias simbólicas del trazado geométrico. En este mapa, el norte se halla hacia la izquierda (adaptado de Navarro Segura 1999, 233).

Sin embargo, algunos años después de la propuesta de Navarro Segura surgieron otras ideas referidas al diseño de la ciudad, que aunque acuerdan con su origen geométrico, difieren en algunos detalles. Por ejemplo, se señala que el mapa del casco histórico no muestra huellas claras de un diseño basado en los radios de una circunferencia. Esto se evidencia a partir del aspecto morfológico de la ciudad, como así también por el trazado de las calles y la ubicación de los edificios antiguos más representativos (Herráiz Sánchez 2007, 61). Por otra parte, comprobada la alineación de tres construcciones antiguas (la torre de La Concepción, la Catedral y la ermita de San Miguel), se plantea la hipótesis de que existe un nuevo eje orientador de la grilla ciudadana. Este sería paralelo al eje de simetría de La Concepción, señalando el horizonte solsticial, y abarcaría toda la ciudad. También se sugiere que la relevancia del eje este-oeste (equinoccial) enfatizado por Navarro Segura no encuentra una adecuada justificación en una ciudad que, supuestamente, está orientada hacia una dirección solsticial. Por último, Herráiz Sánchez sugiere que el ángulo que forma el eje de La Concepción con respecto al punto cardinal este, medida que él estima en aproximadamente 26.5°, representa el ángulo menor de un "triángulo áureo". A partir de esto, se propone que "los contornos de la ciudad, la ubicación de los edificios religiosos y la orientación y dimensiones de las calles principales son el resultado de adaptar a la orografía del lugar los principios geométricos de la Sección Áurea, coincidentes en la latitud de La Laguna con los movimientos solares más relevantes" (Herráiz Sánchez 2007, 85).

De esta forma, el Adelantado habría aplicado a La Laguna el proyecto de ciudad definido por Platón en Las Leyes, su inconcluso y último trabajo filosófico (Navarro Segura 1999, 166). Esto es, se habría relacionado a la nueva fundación con una ciudad ideal, circular como el alma y el universo, imaginada siglos antes por el gran filósofo, propia de una sociedad humana utópica donde se desterraba la deslealtad, sin murallas, formada por círculos concéntricos, y dotada de las estructuras económica, social y política como habían sido sugeridas de acuerdo al antiguo texto (Platón 1943).

## LAS CONSTRUCCIONES RELIGIOSAS EN LA LAGUNA

Desde el momento mismo de la conquista de Canarias, la llegada de la población europea se intensificó, especialmente aquella procedente de las costas de Andalucía occidental. La corriente migratoria que acompañó a los primeros conquistadores comenzó entonces su adaptación en las nuevas tierras. Y sin duda, el origen de estas gentes fue determinante en las diferentes realizaciones arquitectónicas que fueron surgiendo en las poblaciones recién creadas (Corbella Gualupe 2000). Esta nueva arquitectura, de un modo u otro, estuvo siempre ligada a la economía de la Iglesia. Allí donde los colegios eclesiásticos contaron con medios suficientes, optaron por construcciones religiosas dotadas de arcos y formas ojivales. Pero donde las parroquias eran pobres o apenas comenzaban su vida de colectividad, como en el caso de las Canarias, entonces no hubo mucha elección. Este arte religioso debió ser un arte popular, simple y rápido de hacer, y económico. Todas estas eran características que el estilo mudéjar reunía adecuadamente (Fraga González 1977; Rodríguez Estévez 2011).

En Canarias, el barroco abarca aproximadamente desde mediados del siglo XVII hasta las postrimerías del siglo XVIII. Como estilo que se arraiga en las islas, se manifestó en todas las artes y contó con un aporte importante de los creadores locales, cuyos talleres abastecían de cuadros e imágenes a varios sectores de la sociedad. Durante el siglo XVII se consolida entonces la arquitectura tradicional canaria, ya se trate de casas privadas o de construcciones religiosas. Predomina el uso de la madera, abundante en las islas, ya sea en las ventanas exteriores, como en las balaustradas de los patios internos y en los artesonados de los salones. La tradición artesanal de la carpintería de origen mudéjar dejó a su paso obras extraordinarias. Muchos ejemplos se hallan en casas particulares, pero también quedan obras notables en iglesias antiguas, con hermosas creaciones en techos y artesonados, lugares que son de principal interés en un estudio como el que nosotros estamos abordando aquí.

Entre estas construcciones religiosas, podemos mencionar algunas que resultan ser muy representativas. El Real Santuario del Cristo de La Laguna, por ejemplo, adquirió su forma final, dentro de los esquemas propios de la arquitectura tradicional canaria, luego del incendio de 1810 (Fig. 4). En la fachada principal se distinguen dos elementos bien diferenciados. Primero, un paño de cantería que soporta la espadaña, formada por dos cuerpos separados por una estrecha cornisa. Consta de tres arcos de medio punto donde se ubican las campanas y se caracteriza por una decoración barroca de volutas y jarrones ornamentales. En segundo lugar, vemos el lienzo de mampostería donde se ubica la portada principal. Esta está delimitada por un arco de medio punto hecho en cantería roja sobre pilastras, y cuenta con una moldura delgada que discurre bajo el alero.

El Cristo, como se llama comúnmente a esta iglesia, presenta una planta rectangular con una nave única distribuida en tres tramos, divididos por el arco toral que precede al presbiterio y por el arco de medio punto que sostiene la tribuna donde se ubica el órgano. El volumen de la nave y el artesonado austero muestran influencias neoclásicas, dentro de un patrón general de corte barroco.

En el testero se luce un espléndido retablo mayor recubierto de plata repujada que alberga la imagen del Cristo.

Otro ejemplo notable es el convento de Santa Clara de Asís y San Juan Bautista. Aunque construido en el siglo XVI, un incendio acaecido en 1697 destruye gran parte del edificio. Su reconstrucción se emprende poco tiempo más tarde y, ya en 1700, vuelve a abrir sus puertas al culto. Desde el exterior del monasterio destaca el prominente mirador de gusto mudéjar o ajimez, que data de 1717. La iglesia es de nave única, con presbiterio rectangular, al que se accede por un arco de medio punto. Por su valor artístico se destaca la cubierta de la capilla mayor, con una armadura ochavada, también de influencia mudéjar.

De más está decir que los ejemplos abundan, como estos dos que acabamos de enumerar. Ya se trate de la ermita de Nuestra Señora de Gracia, cuya puerta es de madera antigua, apeinazada, con cuarterones tallados, y que presenta un aspecto típico del barroco tardío. O bien, la ermita de San Benito Abad, originariamente de 1554, que posee la típica nave y capilla mayor cuadrilonga, con armadura mudéjar (Corbella Gualupe 2000). O incluso el prominente saledizo o ajimez de estilo mudéjar del convento de Santa Catalina de Siena (Fig. 4). En total, se han contado más de 600 edificios de estilo mudéjar en la ciudad, como dijimos, la mayoría casas particulares, pero también muchas iglesias, casi siempre con detalles característicos en techos y artesonados. El ejemplo quizá más prominente es el de la iglesia de Nuestra Señora de la Concepción, la que fuera el primer templo fundado en la isla tras la conquista. Es una construcción de tres naves, con la central más larga, que incluye el antepresbiterio y el presbiterio, y divididas por arcos de medio punto que descansan sobre columnas toscanas. Constituye un ejemplo icónico de la llamada iglesia mudéjar columnaria. Reflejo de la austeridad de sus materiales y la sencillez de sus líneas, este modelo fue repetido en abundancia por toda Canarias (Rodríguez Reyes 2016).

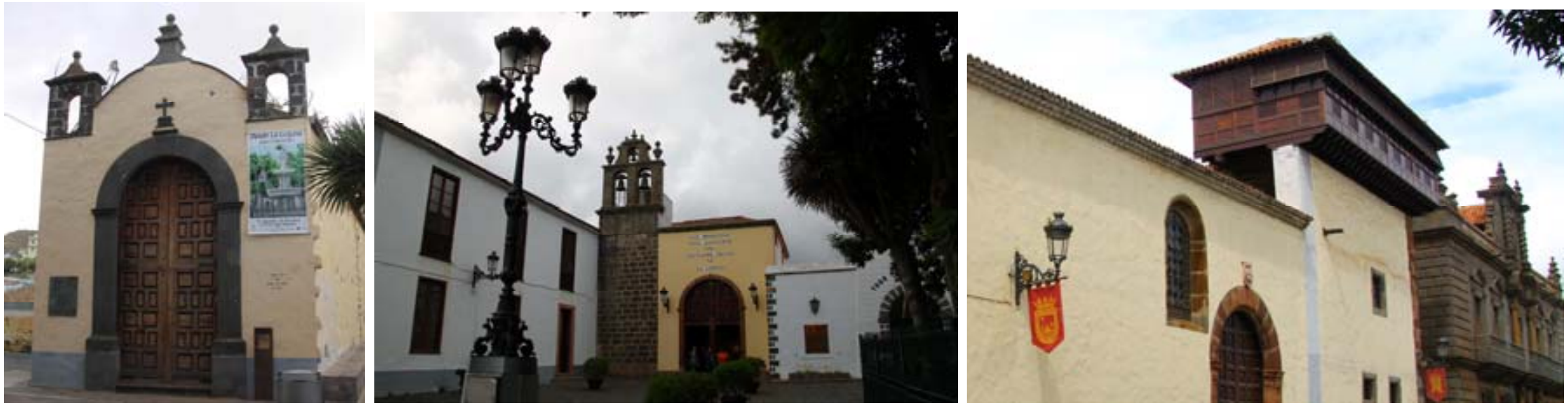

Fig. 4 Tres de las construcciones más antiguas de la ciudad. La ermita de San Miguel de los Ángeles (izquierda) fue fundada en 1506 por el Adelantado Fernández de Lugo, quien expresó su deseo de ser enterrado allí mismo llegado el momento. El Real Santuario del Cristo de La Laguna (al medio) fue fundado en el s. XVI y debió soportar varios problemas edilicios a lo largo de su historia, como consecuencia de inundaciones e incendios. El convento de Santa Catalina de Siena (derecha), antigua mansión del Adelantado en la ciudad, donde se destaca el prominente ajimez que remata la torre, notoria manifestación del mudejarismo en las Canarias. Fotografías del autor.

Pero en este trabajo, más que las características arquitectónicas particulares de las iglesias antiguas, nos interesan sus emplazamientos originales y las orientaciones de sus ejes principales, pues son estos los elementos que nos pueden brindar información sobre la posible intencionalidad de sus constructores y, quizá, sobre el mismo diseño de la ciudad original en la que se fueron emplazando.

## IGLESIAS HISTÓRICAS COMO INDICADORES DE LA PLANIMETRÍA DE LA CIUDAD

Intentaremos aquí presentar otra manera de abordar el estudio del diseño urbano de esta ciudad, con elementos más firmes y concretos que las mencionadas hipótesis -geométricas y algo osadas- de los autores precedentes. Mostraremos que una interpretación más simple de la planimetría actual de la ciudad es posible, y que este plan podría basarse en las orientaciones de sus iglesias (McCluskey 2015). En lo que sigue, y brevemente, resumiremos un estudio exhaustivo que hemos hecho recientemente sobre la orientación del conjunto de iglesias y ermitas laguneras antiguas (Gangui y Belmonte 2017). Se trata de un abordaje arqueoastronómico que nos permitirá llevar luz sobre la constitución de la ciudad en su conjunto y, quizás en un futuro, develar las ideas maestras en el origen de su diseño.

El procedimiento empleado es similar al que presentamos en ocasión del estudio de los templos andinos del norte de Chile (Gangui et al. 2016b). Desde hace años se sabe que la orientación espacial de las iglesias cristianas antiguas es una de las características salientes de su arquitectura. En el continente europeo y en multitud de sitios lejanos que fueron evangelizados, existe una marcada tendencia a orientar los altares de los templos en el rango solar. Es decir, el eje de simetría de una iglesia, desde la puerta principal y en dirección al altar, se orienta hacia aquellos puntos del horizonte por donde sale el Sol en diferentes días del año. Dentro del mismo rango solar, sin embargo, no son infrecuentes las alineaciones en sentido contrario, con el altar a poniente, aunque resultan excepcionales pues no siguen el patrón canónico (Gangui et al. 2014). En nuestro estudio, esta última posibilidad también será tenida en cuenta, pues la orientación a poniente es característica en los tiempos iniciales del Cristianismo y, asimismo, se ha repetido en zonas ya estudiadas del norte del África que son posibles regiones de origen de la población canaria nativa (Esteban et al. 2001; Belmonte et al. 2007).

En nuestro trabajo de campo hemos obtenido datos de las orientaciones precisas de 21 iglesias y ermitas antiguas de La Laguna. Aparte de las coordenadas geográficas en latitud y longitud, medimos los acimutes de los ejes de las construcciones junto con las alturas del horizonte hacia donde apuntan esos ejes. Esto último es necesario, ya que estas dos coordenadas "locales" (acimut y altura) nos permiten conocer exactamente "el punto" en el cielo -en todo el cielo- hacia donde "apunta" cada iglesia. Una vez que tenemos en cuenta la declinación magnética de cada sitio particular (que altera nuestras mediciones de acimut hechas con brújulas de precisión) y la refracción atmosférica (que modifica nuestros valores de la altura, y que es tanto mayor cuanto menor es esa altura medida sobre el horizonte), podemos transformar nuestros datos en valores de la llamada "declinación astronómica". Estos últimos valores no dependen del sitio particular en donde uno se halle, pues la declinación astronómica es una coordenada "ecuatorial", es decir que se mide con respecto al ecuador celeste, y por lo tanto es independiente del lugar en donde se haya hecho la medición.

La idea ahora es comparar estos valores de declinación astronómica de las iglesias con los valores de declinaciones posibles que tiene el Sol a lo largo del año. Como sabemos, debido a la inclinación del eje terrestre con respecto a la eclíptica (el plano de la órbita de la Tierra alrededor del Sol), los valores de la declinación solar varían entre aproximadamente -23.5° y +23.5° (correspondientes a los solsticios de diciembre y de junio, respectivamente). Declinaciones de iglesias fuera de este rango de valores no pueden estar orientadas al Sol. Por ejemplo, una iglesia que se oriente en acimut muy cerca del norte (acimut de 0°) o del sur (acimut de 180°), es decir, cuyo eje se disponga muy cerca de la meridiana del lugar, con toda probabilidad no estará orientada con el Sol, pues -en las latitudes intermedias de los sitios estudiados- el Sol jamás toca el horizonte cerca de los puntos cardinales norte o sur.

Para aquellas iglesias que efectivamente están orientadas dentro del rango solar, es decir que tienen valores de la declinación en el rango (-23.5°, +23.5°), si suponemos que fueron construidas siguiendo al Sol, entonces la declinación de cada iglesia nos indicará la declinación del Sol en el momento de la construcción. Pero como esta última nos define en forma precisa el día del año solar correspondiente (o el par de días, si nos hallamos fuera de los solsticios), entonces a partir de esta coordenada podemos deducir simplemente la fecha del año que privilegia cada iglesia.

Por supuesto, hay varios detalles adicionales que no estamos mencionando, pero que pueden consultarse en las publicaciones ya citadas. Un elemento importante en esta discusión es nuestra suposición de que las iglesias se orientan con el Sol (esa es nuestra hipótesis de partida). Así lo suponemos porque, como mencionamos más arriba, la tradición en la construcción de las iglesias cristianas así lo sugiere.

Todo el análisis de las mediciones puede sintetizarse en un gráfico que muestra nuestros resultados. En la Figura 5 se grafica el número de iglesias con una dada declinación astronómica versus el valor absoluto de la declinación. Tomamos el valor absoluto de esta coordenada pues en este estudio sólo nos interesa considerar la dirección del eje de cada iglesia. Esto se justifica porque, como ya hemos mencionado, existen casos en donde son frecuentes las alineaciones con el altar a poniente. Por otra parte, el número de iglesias con un dado valor de declinación es graficado como "frecuencia relativa normalizada", lo que significa que a los datos calculados se les han sustraído los valores que caracterizan una distribución uniforme de orientaciones. En otras palabras, si los datos de nuestras mediciones fueran compatibles con una distribución uniforme en acimut (es decir, que no existe orientación privilegiada por las iglesias), la curva de la Fig. 5 debería ser horizontal. Por el contrario, se ven algunos picos, y esto muestra que el grupo de iglesias analizado sugiere una dirección particular (hacia un punto del cielo, señalado por la declinación) y, por lo tanto, de acuerdo a nuestras hipótesis, también una fecha especial.

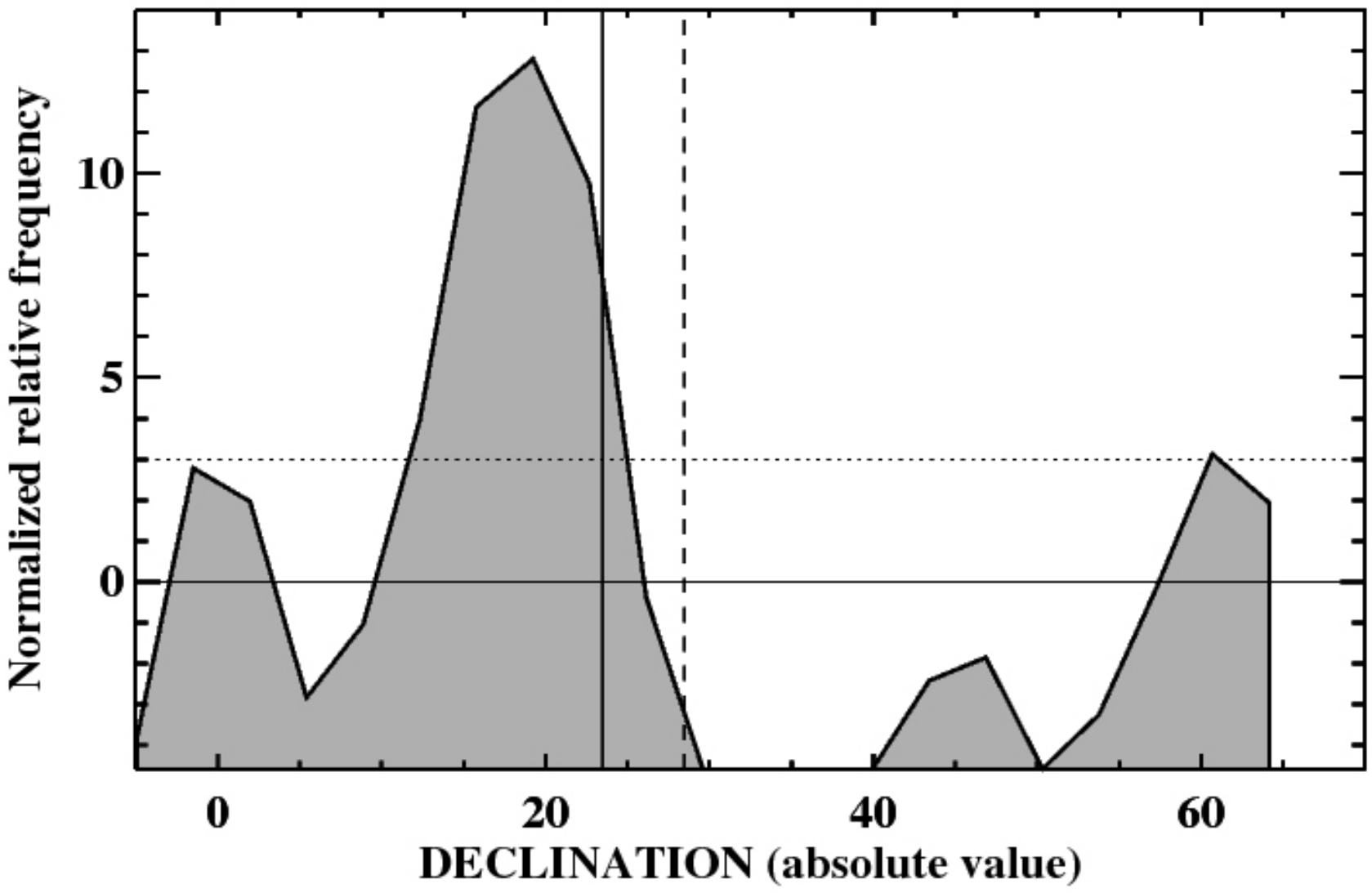

Fig. 5 Histograma de declinaciones en valor absoluto correspondiente a las iglesias y ermitas de La Laguna. Sólo aparece un pico prominente, ubicado en una declinación apenas menor a los 20°. Este pico podría estar asociado con la fiesta católica de San Cristóbal del 25 de julio (o del 10 de julio, ya que ambas fechas son relevantes entre las festividades de la ciudad), pues corresponde a la declinación del Sol en ese día, si nos ubicamos en la época de la fundación de la ciudad. La línea vertical continua representa la declinación correspondiente al valor extremo del Sol en los solsticios, mientras que la línea vertical discontinua indica el valor de declinación extremo correspondiente a la Luna.

La figura nos permite distinguir que existe una cierta "estructura" en las orientaciones de las iglesias, pues aparece un pico notorio que domina el histograma. Este, sin embargo, no es lo suficientemente preciso (angosto) como para seleccionar fehacientemente un dado valor de declinación (y por lo tanto, una fecha precisa). A pesar de esta incertidumbre, en gran parte debida a la dispersión de los datos y a contar con pocas mediciones, pues el número de iglesias de la ciudad en relativamente pequeño, el resultado es sugerente. La tendencia en las iglesias que muestra el gráfico pareciera seleccionar una fecha cercana a la de la fiesta de San Cristóbal de Licia, el santo patrono al cual la ciudad de La Laguna fue originariamente dedicada.

Por otra parte, tanto las orientaciones solsticiales (alrededor de la declinación 23.5°) como las equinocciales (alrededor del cero de declinación) se evidencian subdominantes frente al pico principal. Este resultado muestra que las propuestas de autores anteriores que sugerían ejes geométricos para la ciudad, ya fueran alineados con algún solsticio o con el equinoccio, carecen de sustento si uno ha de creer en las iglesias antiguas como rectoras de la planimetría de la ciudad.

## DISCUSIÓN

En este trabajo estudiamos la orientación espacial de veintiuna iglesias cristianas históricas entre las que existen actualmente en la parte antigua de San Cristóbal de La Laguna, con el fin de proporcionar nuevos conocimientos sobre la planificación de la ciudad. Según la declaración de la UNESCO, La Laguna muestra signos de un intercambio de influencias entre las culturas hispano-portuguesa y americana. Esta característica es evidente no solo en su plan de cuadrícula, sino también en sus iglesias y ermitas, en sus claustros y en la arquitectura civil, que están estrechamente relacionados con las del nuevo continente. La Laguna fue la primera ciudad colonial española no fortificada con una grilla regular, y se cree que su diseño proporcionó el modelo para muchas ciudades coloniales en las Américas. Fruto de la inspiración del Adelantado, Don Alonso Fernández de Lugo, con la gracia y el apoyo de los Reyes Católicos, sobresale en su planificación como una ciudad-territorio, supuestamente construida en un proyecto completo y autónomo como un espacio para la organización de un nuevo orden social (UNESCO 1999). Sin embargo, como también hemos descrito en este documento, las creencias populares actuales sugieren que La Laguna se habría diseñado en acuerdo con ciertas ideas utópicas -propias de una ciudad ideal- inspiradas en antiguos principios filosóficos y geométricos griegos.

En nuestro enfoque del problema elegimos un camino más pragmático para el estudio de la disposición real de la ciudad, mediante el análisis de la orientación exacta de sus iglesias y ermitas antiguas. Pudimos determinar que en La Laguna existen indicios de la existencia de un patrón de orientación definido, y esto indica que las razones para el diseño real de la ciudad podrían ser más simples que lo que habían imaginado autores anteriores. Los datos arqueoastronómicos obtenidos para las iglesias, como se presentan, por ejemplo, en el histograma de declinación de la Figura 5, muestran un patrón representativo que, dentro de ciertos límites impuestos por la limitada muestra de construcciones medidas y el ancho resultante del pico principal, sugiere que existe una fecha privilegiada que es cercana a la fiesta de San Cristóbal de Licia, el santo patrono a quien la ciudad fue originariamente dedicada.

En conclusión, y sobre la base de la discusión anterior, creemos que la planificación de la ciudad, resultante de la distribución espacial de los edificios religiosos estudiados, podría entenderse adecuadamente a partir de principios simples, bien probados, como los empleados usualmente en la investigación arqueoastronómica. Ante estos resultados, creemos que no hay necesidad de proponer hipótesis extrañas, sean estas geométricas o platónicas, sin la adecuada base documental, con el fin de explicar la singular planimetría de la ciudad de San Cristóbal de La Laguna.

*Agradecimientos*:
El autor agradece a Juan Belmonte y sus colaboradores, miembros del Instituto de Astrofísica de Canarias, por la invitación que permitió llevar a cabo el trabajo de campo, y por las intensas discusiones posteriores sobre los resultados. También agradece el apoyo financiero del Conicet y de la Universidad de Buenos Aires.

BIBLIOGRAFÍA